\documentstyle[preprint,aps,psfig]{revtex}
\begin{document}
\draft
\title{Interpenetration of two 
chains different in sizes: Some Exact Results}
\author{Sanjay Kumar$^\ast$} 
\address{Department of Physics, Banaras Hindu University \\
Varanasi 221 005, India, and \\ 
 {$*$}Institute For Theoretical Physics \\
Zulpicher Street 77 \\
University of Koln, D-50937, Koln, Germany}
\date{} 
\maketitle
\begin{abstract}
A model of two interacting polymer chains has been proposed to study the effect of
 penetration of one chain in to the other. We show that small chain penetrates
 more in comparison to the long chain. We also find a condition in which 
both chains cannot grow on their  own (or polymerize) but can grow (polymerize) 
in zipped form.
\end{abstract}
\vspace {1in}

In recent papers [1-3] it has been shown that the lattice model of 
self-attracting-self-avoiding walks (SASAWs) [4-5 ] may be used to study 
the critical behavior of two chemically different interacting polymer 
chains in a solution which may be of different types ( good or poor) for
 different chain. We reported transition from zipped state to segregated 
state by varying the solvent quality, the temperature and the interaction 
between monomers of inter-chain and intra chain. The model proposed by us 
takes in to account of the physical condition that the interaction between 
monomers is repulsive at short distance and attractive at large distance.
 The non crossing constraint represents the repulsion, and the attraction 
between monomers occupying the neighboring lattice site is due to the 
attractive part of the interaction. This model is termed as two-interacting 
walks (TIWs) and is solved exactly for truncated n-simplex lattice 
(n=4,5 and 6) using real space renormalization group calculation. The mean 
number of monomers $M$ of one chain in contact with the other chain at the 
vicinity of transition point is expected to behave as [2].
\begin{equation}
M \sim N ^ y
\end{equation}

Where $ N$ is the total number of monomers in one chain and $y$ is 
the contact exponent.

The aim of this letter is to study the influence of chain length distribution. 
This problem has got considerable attention in past years from experimental 
and theoretical point of view. Experiments have been done with mixture of 
polystyrene [10-11] and poly-${\alpha}$-methylstyrene [12] and the results 
were interpreted by various two parameter perturbation theories assuming 
that polymer coil behaves like the hard sphere. Recent theoretical studies 
have been devoted to this problem [13-15] and renormalization group analysis 
showed that chains  small in sizes penetrate more compare to the long chain. 
Deviation to the hard sphere model has already been seen by the 
experimentalists [10-12] and their argument also lead to the conclusion 
that the polymers of lower molecular weights penetrate more [10-11]. 
In another experiment [16] several mixtures of same chemical species 
with very different sizes, confirms the prediction.

To study the interpenetration of small chains, we consider 
a grand canonical ensemble of polymers. 
By grand canonical we mean,of course, not the number of chains are 
fluctuating but the number of monomers in the chain are fluctuating i.e. 
the chain length is fluctuating. The generating function of our interest is
\begin{equation}
\Omega(x_{1},x_{2},x_{3},u_{1},u_{2},u_{3})  = 
\sum_{all walks}(x_{1}^{N_{1}}
u_{1}^{R_{1}})(x_{2}^{N_{2}}u_{2}^{R_{2}})(x_{3}^{N_{3}}u_{3}^{R_{3}})
\end{equation}
where $N_{1}(N_{2})$ is the number of steps (monomers) in the polymer 
$P_{1}(P_{2})$ and $x_{1}(x_{2})$ denotes the fugacity weight attached to 
each step of polymer $P_{1}(P_{2})$ respectively. Here $u_{1}(u_{2})$ 
represents the Boltzmann factor associated with the attractive interaction 
between monomers of polymer $P_{1}(P_{2})$
. $x_{3}$ and $u_{3}$ denote, respectively, the fugacity weight attached 
with zipped walks and the Boltzmann factor associated with the attractive 
interaction between monomers of $ P_{1}$ and $P_{2}$. 
$N_{3}$ is the total number of zipped steps. $R_{1}(R_{2})$ is the number of
 monomers of $P_{1}(P_{2})$ occupying the nearest neighboring sites of the 
lattice and $R_{3}$ is the number of monomers of different chains occupying 
the nearest neighbor lattice points. The complete phase diagram with these 
fugacities is given in [2].

For present study we consider  situations in which $x_{1}(x_{2}) 
< x_{1}^{c}(x_{2}^{c})$ and or $u_{1}(u_{2}) < u_{1}^{c}(u_{2}^{c})$ 
such that long chains are exponentially suppressed and the ensemble is 
dominated by the short chains. We shall use this condition in our 
further study. The generating function expressed in terms of Eq.(2) 
for $4$-Simplex lattice[1-5]
can be represented in terms of restricted partition functions shown in 
Fig. 1 and their corresponding recursion relations are given below 

\begin{eqnarray}
A_{r+1}=A^{2}+2A^{3}+2A^{4}+4A^{3}B+6A^{2}B^{2}
\end{eqnarray}

\begin{eqnarray}
B_{r+1}=A^{4}+4A^{3}B+22B^{4}
\end{eqnarray}

\begin{eqnarray}
C_{r+1}=C^{2}+2C^{3}+2C^{4}+4C^{3}D+6C^{2}D^{2}
\end{eqnarray}

\begin{eqnarray}
D_{r+1}=C^{4}+4C^{3}D+22D^{4}
\end{eqnarray}

\begin{eqnarray}
E_{r+1}=(AC)^{2}+2ACE(A+C)+2E^{4}+6E^{2}(B^{2}+D^{2})+4E^{3}(B+D)
\end{eqnarray}
Here and below we adopt a notational simplification in which the 
index $r$ is dropped from the right hand side of the recursion relations. 
It may be emphasized here that the recursion relations written above 
are exact for the model defined above. The recursion relations for $A$ 
and $ B$ are  independent of $ C,D $ and $ E$ ( and $C$ and $D$ are 
independent of $A, B$ and $E$) is 
the consequence of the definition of the model. This means that the 
model defined above take care of the criticality of one chain in dilute solution 
does not 
get affected in presence of other chain. 
However, The effect of a polymer chain on other  has been taken  through
intercation mediated by configuration 
$E$ . Since the interaction between inter chain and or intra chain   
is  restricted to bonds within a first order unit of the 
fractal lattice, $u_{1},u_{2}$ and $u_{3} $ do not appear explicitly in the 
recursion relations.They appear only in the initial values [4-5] given 
below:
\begin{eqnarray}
&   A_{1}=x_{1}^{2}+2x_{1}^{3}u_{1}+2x_{1}^{4}u_{1}^{3}
\end{eqnarray}      
\begin{eqnarray}
&   B_{1}=x_{1}^{4}u_{1}^{4}
\end{eqnarray} 
\begin{eqnarray}
&   C_{1}=x_{2}^{2}+2x_{2}^{3}u_{2}+2x_{2}^{4}u_{2}^{3}
\end{eqnarray}
\begin{eqnarray}
&   D_{1}=x_{2}^{4}u_{2}^{4}
\end{eqnarray}

\begin{eqnarray}
&   E_{1}=(x_{3}u_{3})^{4}
\end{eqnarray}
Here index 1 and 2 on the right hand side of Eq.(8-11) correspond to 
chain $P_{1}$ and $P_{2}$ respectively and  $x_{3}=(x_{1}x_{2})^{1/2}$ is the 
fugacity of zipped walk. The fixed points 
corresponding to different configurations of polymers in the asymptotic 
limit are found by solving Eq.(3-4) for polymer chain $P_{1}$ and 
Eq.(5-6) for polymer chain $P_{2}$ respectively. A complete phase 
diagram of Eq.(3-4) or Eq.(5-6) is given in [4-5]. The state 
of polymer chain $P_{1}(P_{2})$ depends on the quality of the solvent and 
on the temperature and can therefore be any of three states [17]; Swollen, 
Compact globule and at $\Theta$ point described in the asymptotic 
limit by the fixed points $ (A^{*},B^{*}) $ = ( 0.4294, 0.0498), 
(0.0,$ 22^{-1/3}) $and (1/3,1/3) respectively. The fixed point 
corresponding to the swollen state is reached for all values of $u < 
u_{c}$  at $x = x_{c}(u)$. The end to end distance for a chain 
of $N_{1}$ monomers of $P_{1}$  in this state varies as $ 
N_{1}^{\nu_{1}}$ with $\nu_{1}$ =0.6740 [4-5] and connectivity 
constant $\mu$ =$1/x_{c}$ is found to be 1.5474. The fixed point 
corresponding to the compact globule state is reached for all values 
of $ u > u_{c}$ at $ x_{1}=x_{c}(u) $. At $u_{1}$ = 3.31607.. and 
$ x_{c}$ = 0.22913.. the system is found to be at its tri-critical 
point or $\theta $ - point. 

We consider a situation in which either one or both chains have fugacity less 
than their critical value. This represents a state in which a long chain can 
not exist independently and system is dominated by small chains. It
 is indicated by ${\bf O}$ state which 
corresponds to the fixed point $(A^{*},B^{*})$ and or $(C^{*}D^{*})$ 
= (0.0, 0.0). The four independent combinations found in these cases 
are achieved by considering following conditions.
${\bf SO}$ represents the situation in which one of the polymer
chain is in the swollen state (indicated by${\bf S}$) with the condition
 $(x_{1}=x_{1}^{c}(u_{1})$ with $u_{1} 
< u_{1}^{c}$ while the other chain is in ${\bf O}$ state with the condition
 $x_{2} < x_{2}^{c}(u_{2})$ with $u_{2} < u_{2}^{c}$). 
${\bf TO} $ corresponds to the situation in which one of the 
chain is at ${\theta -point}$ with the condition $x_{1}=x_{1}^{c}(u
_{1})$ with $u_{1}$ = $u_{1}^{c}$ and second chain is in ${\bf O}$ state
 ($x_{2} < x_{2}^{c}(u_{2})$ with $u_{2} 
< u_{2}^{c})$). ${\bf CO} $ represents that one of the polymer chain
is in collapsed state with the condition ($ x_{1}=x_{1}^{c}(u
_{1})$ with $u_{1}  > u_{1}^{c}$ while the second one is in ${\bf O}$ state.
 ${\bf OO}$ indicates the situation in which both the chains are 
at below of their criticallity ($ x_{1} < x_{1}^{c}(u
_{1})$ with $u_{1} < u_{1}^{c}$ and $x_{2} < x_{2}^{c}(u_{2})$ with $u_{2} 
< u_{2}^{c})$.

The four non trivial fixed points reached by Eq. (7) using Eqs(3-6) 
are given in table 1. These fixed points are reached by the system 
when $x_{3}$ and $u_{3}$ lie on the curve drawn in Fig. 2. 
We find that the value of $x_{3}$ as a function of $u_{3}$ depends on the 
states of individual chain. Linearization of Eq.(7) about these fixed 
points lead to the eigenvalue $\lambda_i > 1 $ which gives contact exponent
which we list in table 1 
by the relation given below[1-3]
\begin{equation}
y=ln(\lambda_{i})/ln(\lambda_{b})
\end{equation}
where $\lambda_{b}$ is the largest eigenvalue of the system. The curves 
between $u_{3}$ and $x_{3}$ are continuous without any multicritical point.

For five simplex lattice the polymer chain [5] always remain in the swollen
 state for all values of intra-chain interaction  and hence we have only 
two possible combinations of chains i.e. ${\bf SO}$ and ${\bf OO}$. 
The results obtained by these combinations are given in table 2. 
However for 6-simplex lattice we do find the similar behavior as of 
4- simplex lattice. The fixed points corresponding to ${\bf SO, CO,TO}$ 
and ${\bf OO}$ and corresponding contact exponents are given in table 3.
The recursion relations and coressponding partition functions may be taken
from[4-5] for$5$ and $6$-simplex lattices.
It is interesting to compare our results with experimental observations. 
We find an unbinding transition  from a configuration zipped states of 
the chain to a state where the chains are segregated. The unbinding transition 
point  is a multicritical point. At this point the monomers of a chain in 
contact with the monomers of other chain scales with a exponent $y$ and 
is called  the contact exponent.

The value of $y$ listed in table 1 for a single chain surrounded by small 
chains, exhibits an interesting behavior. As is evident from table 1, 2 
and 3 that when one chain is in $S$- state and other is at ${\bf O}$ state the 
exponent $y$ is found to be greater than one. It may be noted here that 
${\bf O}$ state corresponds to the situation in which long chain does 
not exist. and system is dominated by small chain. It may be thought as 
instead of one chain combining all monomer dissolved in the solution, we may 
have several small chains. Therefore the monomers of swollen chain will be 
in contact with several monomers of the small chains, which make the exponent
 larger than one. This may not be the case when one of the chains is in compact 
globule or at tricritical state and the other at ${\bf O}$ state. 
Most of the monomers of the first chain are occupied by its own 
monomers and therefore making contact with other monomers of small chain less 
probable. Hence the contribution will come mostly from the surface of the 
collapsed polymer. Recent computer simulation with melt polymer have similar 
behavior where interpenetration is found to be the shape dependent[9]. 
However, when the polymer chain is at its tri-critical, we find that contact 
exponent is larger than the compact one. Because in this case, small chain can 
penetrate.It will be interesting to note that when two chains are in swollen 
state[1] the contact exponent was found to be 1 indicating that one monomer 
of a chain is in contact with a monomer of other chain and is in zipped state.

The other interesting feature which is obvious from table 1,2 and 3 is that in 
a solution in which none of the chain can polymerize on its own (i.e. the 
chains are individually are in ${\bf O}$ state but can polymerize in zipped 
form due to unlike monomer-monomer interaction. We note that this zipped 
state configuration of two chains in all cases is in compact phase as is 
evident from eigenvalue listed in table 1,2 and 3. Apart from this in all 
other cases when chains are in either T or C state, the value of y is found 
to be less than 1. 

We thank Yashwant Singh and Deepak Dhar for many useful discussions on the subject. We also thank INSA-DFG for financial assistance.
\newpage
\begin{table}
\caption{{\it Fixed point of $ E$ state, eigenvalues and contact exponent for 
4-simplex lattice}}
\begin{tabular}{|l|l|l|l|l|} 
State of chain &$ E_{i}^{*}$ &$ \lambda_{i}$ & $\lambda_{b}$ &$ y$ \\
\tableline
SO & 0.7857 & 3.8617 & 2.7965 & 1.3138 \\
CO & 0.5036 & 2.8769 & 4.000 & 0.7622 \\
TO & 0.5231 & 2.9371 & 3.7037 & 0.8228 \\
OO & 0.7937 & 4.00 &------ &- \\ 
\end{tabular}
\end{table}

\begin{table}
\caption{{\it Fixed point of $ E$  state, eigenvalues and contact exponent for 
5-simplex lattice}}
\begin{center}
\begin{tabular}{|l|l|l|l|l|} 
State of chain & $E_{i}^{*}$ & $\lambda_{i}$ &$ \lambda_{b}$ &$ y$ \\ 
\tableline 
SO & 0.4516 & 3.9809 & 3.1319 & 1.2101 \\
OO & 0.5694 & 4.6308 &- &- \\ 
\end{tabular}
\end{center}
\end{table}

\begin{table}
\caption{{\it Fixed point of $G$,$ H$ and $I$[4-5] states,eigenvalues
 and contact exponent for 
6-simplex lattice}}
\begin{center}
\begin{tabular}{|l|l|l|l|l|l|l|} 
State of chain &$ G_{i}^{*}$ &$ H_{i}^{*}$ &$ I_{i}^{*}$ &$ \lambda_{i}$ &$ \lambda_{b}$ &$ y$ \\ 
\tableline 
SO & ~0.0 & 0.1039 & 0.1049 & 5.9929 & 3.4965 & 1.4304 \\
TO & 0.9628 & 0.0795 & 0.1048 & 5.3050 & 5.4492 & 0.9341 \\
CO & 0.0    & 0.0711 & 0.1050 & 5.057 & 6.00 & 0.9046 \\
OO & 0.0    & 0.1042 & 0.1042 & 6.0 &- &- \\
\end{tabular}
\end{center}
\end{table}


\begin{figure}
\caption{diagrams representing the five restricted partition functions for two chains
( indicated by black$ (P_{1})$ and white$ (P_{2})$ lines on 4-simplex lattice.
}
\end{figure}

\begin{figure}
\caption{ critical fugacity $x_{3}$ for entanglement of two chains as a function of
$ {u_{3}} $ in a situation where one or both chains can not grow on their own. 
Results are shown here for$ CO$ and $OO$ states of individual chains. Other configurations
SO and TO have values in between these two curves. The letter$ O$ indicates a condition in which 
long chain can not grow and system is dominated by small chains.
}
\end{figure}
\end {document}
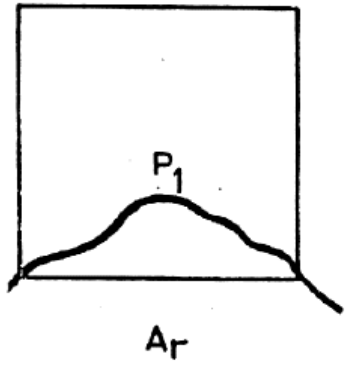
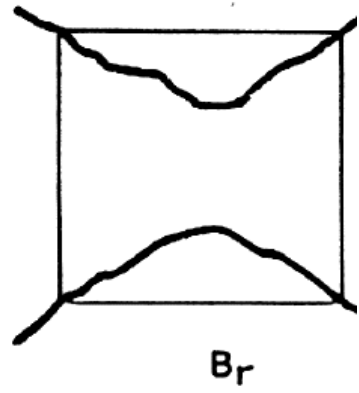
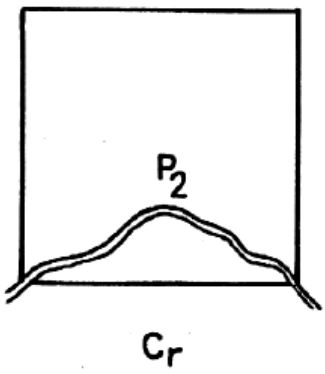
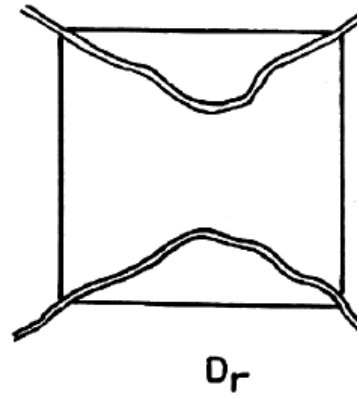
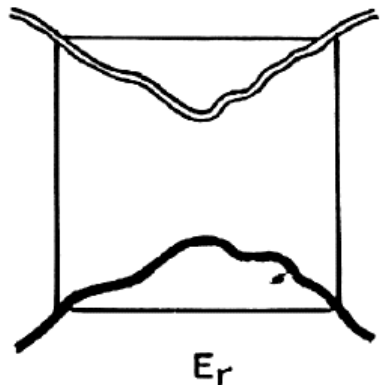
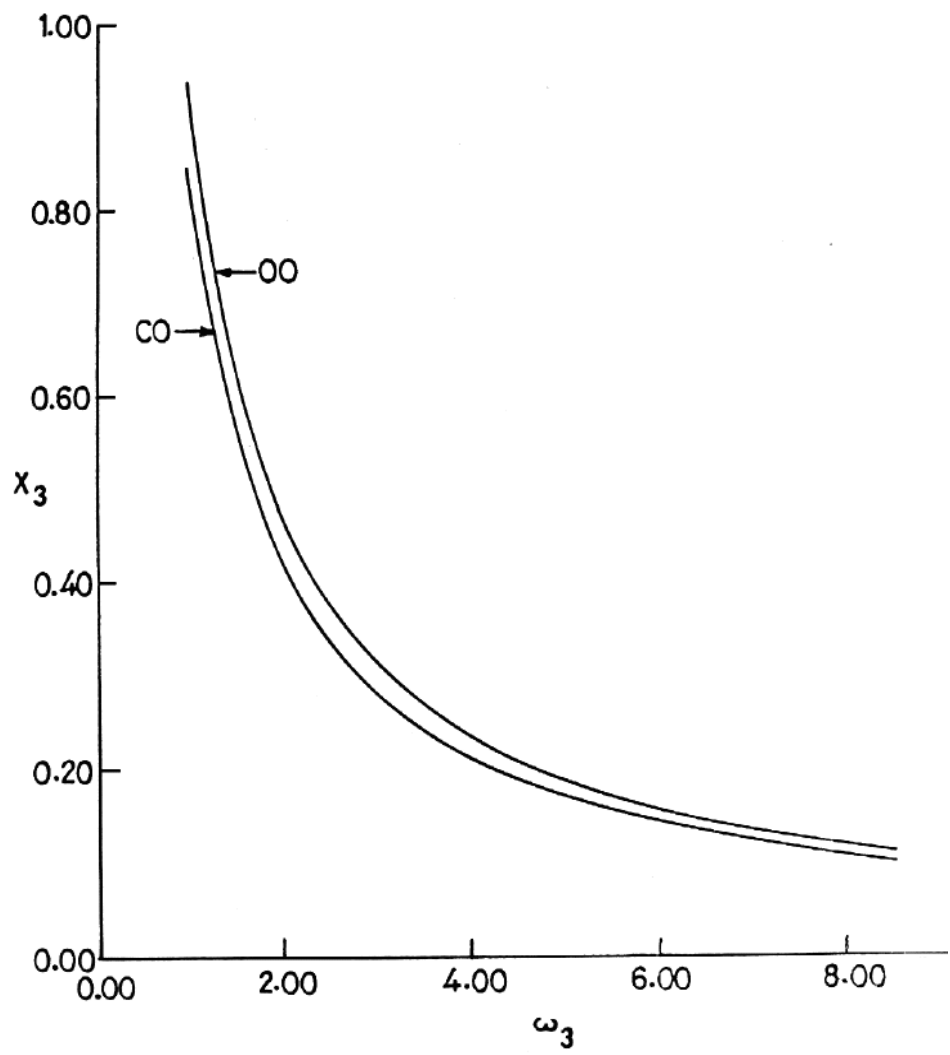
